\documentclass[a4paper,10pt,twocolumn]{article}

\usepackage[super,sort&compress]{natbib}
\usepackage[noblocks]{authblk}
\usepackage{amssymb}
\usepackage{stmaryrd}
\usepackage{subfigure}
\usepackage{amsmath}
\usepackage{pifont}
\usepackage{color}
\usepackage{colortbl}
\usepackage{color}
\usepackage{graphicx}
\usepackage{multicol}
\usepackage{url}
\usepackage{abstract}
\immediate\write18{texcount \jobname.tex -out=\jobname.sum}
\usepackage{verbatim}
\newcommand\wordcount{\verbatiminput{\jobname.sum}}

\begin{document}

\title{Revealing latent factors of temporal networks for mesoscale intervention in epidemic spread}

\author[1]{L. Gauvin}
\author[1]{A. Panisson}
\author[1,2]{A. Barrat}
\author[1]{C. Cattuto}
\affil[1]{\small Data Science Lab, ISI Foundation, Torino, Italy}
\affil[2]{\small Aix Marseille Universit\'e, Universit\'e de Toulon, CNRS, CPT, UMR 7332, 13288 Marseille, France}


%
%

\twocolumn[
  \maketitle
  \begin{onecolabstract}
\textbf{
The customary perspective to reason about epidemic mitigation in temporal networks
hinges on the identification of nodes with specific features or network roles.
The ensuing individual-based control strategies, however,
are difficult to carry out in practice
and ignore important correlations between topological and temporal patterns.
Here we adopt a mesoscopic perspective and present a principled framework
to identify collective features at multiple scales
and rank their importance for epidemic spread.
We use tensor decomposition techniques to build an additive representation
of a temporal network in terms of mesostructures,
such as cohesive clusters and temporally-localized mixing patterns.
This representation allows 
to determine the impact of individual mesostructures on epidemic spread
and to assess the effect of targeted interventions
that remove chosen structures.
We illustrate this approach using high-resolution social network data
on face-to-face interactions in a school
and show that our method affords the design of effective mesoscale interventions.
}
  \end{onecolabstract}
]

%
Many natural and artificial systems are adequately represented
in terms of interaction networks between their components~\cite{vespignani2012modelling}.
The network representations has led to key insights on the relation
between the structure of these systems and the dynamics of diverse processes that take place over the network,
such as information diffusion, epidemic spread,
and much more~\cite{barrat2008dynamical,onnela2007structure}.
In recent years, the availability of time-resolved network data has pushed network science
beyond the static graph representation and has prompted new research on understanding and modeling
time-varying graphs, commonly referred to as temporal networks~\cite{Review-TNW-Holme}.
This has spurred an intense activity on studying how dynamical processes are affected
by the temporal evolution of  the network on which they unfold,
focusing on heterogeneous distributions of inter-event times (``burstiness''),
heterogeneous activity distributions, causality constraints, etc.~\cite{Vazquez:2007,karsai-2011-83,Perra2012,gauvin2013activity,Rocha:2013,masuda:2013,Takaguchi:2013,Horvath:2014,Holme:2014,Scholtes:2014,Backlund:2014,Perotti:2014}.
High-resolution social network data~\cite{eagle2006reality,salathe2010high,Cattuto2010,Lehmann2014},
in particular, have opened the way to richer individual-based models
of epidemic spread~\cite{salathe2010high,stehle2011simulation,Lee:2012,Gemmetto2014}
and have raised new questions on designing control strategies for epidemic processes
in temporal networks~\cite{Lee:2012,starnini2013immunization,Liu2014,bajardi2011}.
However, micro-interventions that operate on individual nodes, such as targeted vaccination,
are difficult to implement, both for lack of high-resolution data in general cases
and because action at the level of individuals is subjected to many informational and operational constraints.
Therefore, it is natural to consider interventions that target important collective patterns and structural units
of temporal networks. Since network communities, i.e., cohesive clusters, have long been recognized
as key structural units in the architecture of complex networks~\cite{Fortunato2010},
it is natural to think of interventions at the community level~\cite{salathe2010dynamics,Gemmetto2014}. 
Time-varying networks, however, can exhibit a richer range of activity/connectivity patterns than static networks.
Defining and detecting temporal network structures at the intermediate (``meso'') scale,
as well as understanding their relevance for dynamical processes occurring on the networks,
remains a largely open question with important applications.
Early work on the mesoscale structure of complex networks
has mostly focused on static networks~\cite{Focus_mesoscale} 
and on community detection in temporal networks~\cite{Mucha2010}.
Recently, latent factor analysis techniques have been used
to simultaneously detect topological and temporal activity patterns
of time-varying networks~\cite{gauvin2014detecting}.

Here we tackle the problem of designing interventions that selectively target the mesoscopic structure
of a temporal network, with the goal of controlling a dynamical process such as epidemic spread.
This requires both detecting mesostructures and ranking them by their importance for a specific dynamical process.
To detect mesostructures we build on the work of Ref.~\citenum{gauvin2014detecting}
and use tensor decomposition techniques~\cite{Kolda2009}
to represent a temporal network as an additive superposition of mesostructures,
each of which can encode complex structural and temporal correlations.
To assess the importance of individual mesostructures,
we take advantage of the additivity of the tensor decomposition,
which allows us to dissect the original network and reassemble altered temporal networks
where chosen mesostructures are selectively removed.
By comparing the dynamics of a given process on the original network and on the altered ones
we can then determine the specific role of individual mesostructures
and we can rank the effectiveness of interventions that target those structures for removal.

We illustrate our approach using high-resolution social network data on face-to-face interactions in a school setting
and investigating the importance of individual mesostructures for simple epidemic processes.
Schools are actually an interesting context for models of epidemic spread,
as they are thought to play an important role in the community spread
of infectious diseases~\cite{Longini:1982,Viboud:2004,chao2010school,lee2010simulating,Baguelin:2013}. 
Macro-scale interventions such as school closure are considered a viable strategy
for epidemic mitigation~\cite{Cauchemez2009473} but come with steep socio-economic costs~\cite{Brown,Dalton:2008},
thus calling for more targeted approaches.
We show that our approach identifies effective strategies for epidemic mitigation
that involve the targeted removal of individual mesostructures.
Remarkably, we find that the most important mesostructures for epidemic spread
correspond neither to the strongest structures nor to cohesive structures
that can be easily detected by community detection methods.

\section*{\normalsize Mesoscale structure of temporal networks}

\subsection*{\small Latent factor analysis of temporal networks}
We first summarize a general framework that we introduced in ref.~\citenum{gauvin2014detecting}
to expose the mesoscale structure of a temporal network.
The approach combines a tensor representation of a temporal network~\cite{DeDomenico2013}
and a dimensionality reduction technique based on tensor decomposition~\cite{Kolda2009}.

We consider an undirected and unweighted temporal network,
hence the state of the network at time $t$ can be represented
by a binary-valued adjacency matrix $\textbf{M}(t) \in \mathbb{R}^{N \times N}$,
where the matrix entry $M_{ij}(t)$ indicates the status of link $i$-$j$ at time $t$,
and $N$ is the number of nodes of the network.
The matrices describing the state of the network at different times
can be combined into a 3-mode tensor, $\mathcal T \in \mathbb{R}^{N \times N \times L}$,
where the first two dimensions are the customary node indices of the adjacency matrix
and the third dimension is a temporal index.
$L$ is the number of network snapshots, each for a different point in time.
To detect structures, i.e., correlated activity/connectivity patterns of the temporal network,
we set up a latent factor analysis based on non-negative tensor factorization~\cite{gauvin2014detecting}:
The central idea is to approximate the tensor $\mathcal{T}$
with a sum $\tilde{\mathcal{T}}$ of rank-1 non-negative tensors (see Methods):
\begin{equation}
\mathcal{T} \, \simeq \, \tilde{\mathcal{T}} = \sum_{r=1}^R \, \mathcal{S}_r \  .
\label{eq:decomp}
\end{equation}
with $R < \min{\{N,S\}}$.
This decomposition allows to describe the original temporal network
as a purely additive superposition of a chosen number of activity/connectivity patterns,
encoded by the individual tensor components.
Each component $\mathcal{S}_r$ is regarded as a mesoscale structure of the temporal network
and can encode cohesive clusters (i.e., communities),
temporally-localized mixing patterns, and more.
Such a parts-based representation is a general and powerful feature
of tensor decompositions with non-negativity constraints~\cite{lee1999learning}
and greatly enhances the interpretability of individual components.
A similar approach based on non-negative matrix factorization
has been proposed to detect overlapping communities in static networks~\cite{Yang:2013}.

The additive decomposition of Eq.~\ref{eq:decomp} also suggests a natural way
to investigate the role played by individual mesoscale structures,
both on the overall network architecture and on the dynamics of processes
occurring over the network, e.g., epidemic spread.
In particular, we can assess the importance of individual mesostructures
for dynamical processes unfolding over the temporal network
by selectively removing those structures: We can start from the decomposition
of Eq.~\ref{eq:decomp}, remove a chosen term (i.e., a chosen structure),
and sum all the other terms, obtaining a modified temporal network
where the chosen structure has been selectively erased.
We can then simulate a given dynamical process over the
original temporal network and over the modified temporal network,
and compare the dynamics or the outcome of the process,
hence learning about the impact of the removed structure.

\subsection*{\small Case study}
\label{sec:case_structure}
As a case study we consider an empirical temporal network of human face-to-face interactions
in a primary school~\cite{Stehle2011}, measured by using wearable sensors~\cite{Cattuto2010}.
The dataset describes the interactions (or ``contacts'') between $231$ children aged $6$ to $12$
and $10$ teachers, organized in $10$ classes.
It comprises the contacts that occurred at the school premises
during two consecutive days in October 2009, from 8:30am to 5:15pm of each day.
We chose this dataset because it is publicly available~\cite{Stehle2011,Gemmetto2014}
and provides not only a high-resolution temporal network,
but also ground truth information on communities, namely, class attributes for all nodes.
Moreover, the school schedule is known and can be used
to understand the activity patterns that mix multiple classes,
as done in Ref.~\citenum{gauvin2014detecting}.
In the following this will allow us to relate the effect
of removing an individual component
to specific knowledge about the social behavior associated with that component.

The school data are naturally represented as a temporal network,
where nodes are individuals and links are interactions between pairs of nodes.
The networks snapshots of our dataset are recorded every $20$ seconds,
however the school activity is scheduled on a coarser temporal scale.
Hence, we aggregate the data over consecutive time intervals of $15$ minutes
to obtain unweighted network snapshots: for every $15$-minute interval, 
a link is drawn between two nodes if the corresponding individuals
had at least one contact during that interval.
This yields a temporal network represented by binary-valued valued tensor ${\mathcal{T}}$
with $N=241$ and $L=70$.

\begin{figure*}[!ht]
\centering
\includegraphics[width=\textwidth, keepaspectratio]{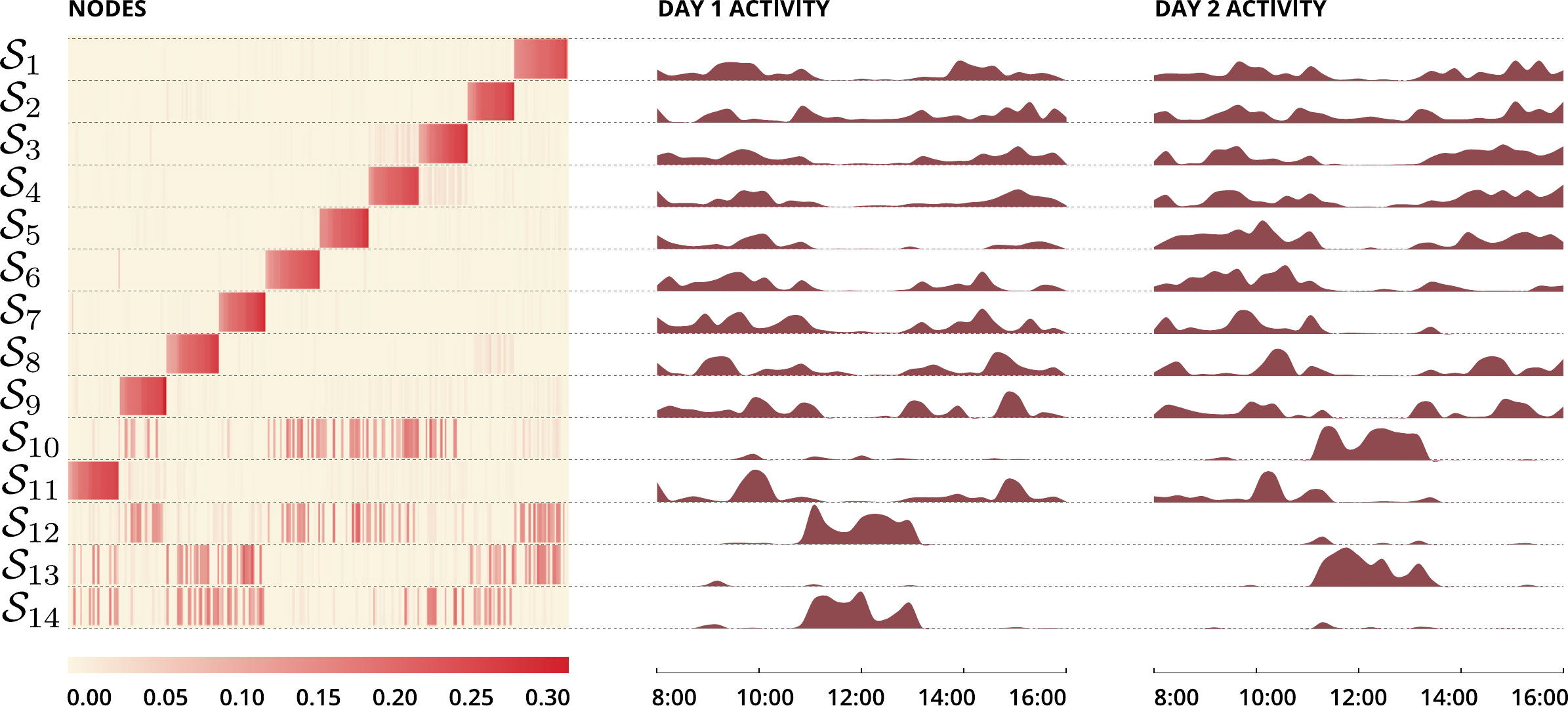}
\caption{
\textbf{Components obtained by tensor factorization of the school temporal network.}
Each row is associated with one component
$\mathcal{S}_r = \mathbf{a_r} \, \circ \, \mathbf{a_r} \, \circ \, \mathbf{c_r}$
of the tensor decomposition $\tilde{\mathcal{T}}$ in Eq.~\protect{\ref{eq:approx}} (see Methods), with $R=14$.
Components are ordered top to bottom according to their Euclidean norm.
Left-hand color-coded matrix: Node weights for each component.
Each row corresponds to a vector $\mathbf{a_r}$
of the decomposition, 
normalized to unity in Euclidean norm for clarity of visualization.
Nodes are ordered to expose the block structure of the matrix due to school classes.
Right-hand plots: Component activity as a function of time over the 2-day span of the dataset.
Each row correspond to a vector $\mathbf{c_r}$, normalized to unity in Euclidean norm.
Components $\mathcal{S}_1$ to $\mathcal{S}_9$ and component $\mathcal{S}_{11}$
are school classes, while components $\mathcal{S}_{10}$, $\mathcal{S}_{12}$, $\mathcal{S}_{13}$
and $\mathcal{S}_{14}$ mix individuals from different classes and are active during lunch breaks.
\label{fig:component}
}
\end{figure*}

Following ref.~\citenum{gauvin2014detecting},
we carry out the non-negative tensor factorization of Eq.~\ref{eq:decomp}
and approximate the tensor $\mathcal{T}$ with the sum $\tilde{\mathcal{T}}$ of $R=14$ components (see 
Methods and Supplementary Information):
the results of the decomposition are reported in Fig.~\ref{fig:component}.
Components $\mathcal{S}_1$ through $\mathcal{S}_9$ and component $\mathcal{S}_{11}$
correspond to the $10$ school classes of the school. They can be validated using the know class
attributes for students and teachers, and they can also be detected to some extent
using community detection algorithms on the temporally-aggregated network~\cite{gauvin2014detecting}.
The components $\mathcal{S}_{10}$, $\mathcal{S}_{12}$, $\mathcal{S}_{13}$
and $\mathcal{S}_{14}$, however, mix individuals from different classes
and -- contrary to the above class components -- they are only active during lunch breaks.
Such components cannot be found by community detection of the temporally-aggregated networks
as they correspond to temporally-localized mixing patterns of the temporal network,
due to mixing of students at lunch time. Tensor decomposition, however, can naturally
detect these patterns as well as the customary cohesive network communities.

We remark that whereas the original temporal network $\mathcal{T}$ is unweighted,
the approximating network $\tilde{\mathcal{T}}$ is in general weighted (see Methods).
For any given network snapshot, corresponding to a $15$-minute time interval,
we will interpret the weight of the edge between two nodes as the cumulated time that those nodes
spent in contact over that $15$-minute interval.
In Table~\ref{table:weights} we report the fraction of total tensor weight corresponding
to each of the $R=14$ components, $\Vert \mathcal{S}_r \Vert \, / \, \Vert \tilde{\mathcal{T}} \Vert$.
\begin{table*}[]
\centering
\begin{tabular}{|c|c|c|c|c|c|c|c|c|c|c|c|c|c|c|c}
\hline
r  & 1 & 2 & 3 & 4 & 5 & 6 & 7 \\
\hline
weight fraction & 11.3\% & 8.6\% & 8.8\% & 7.1\% & 8.3\% & 6.9\% & 5.7\% \\ 
\hline
\hline
r & 8 & 9 & 10 & 11 & 12 & 13 & 14 \\
\hline
weight fraction & 6.8\% & 7.5\% & 7.5\% & 4.8\% & 6.8\% & 5.7\% & 6\% \\
\hline
\end{tabular}
\caption{
\textbf{Component weight.}
Fraction of total tensor weight corresponding to each term $\mathcal{S}_r$ of the decomposition $\tilde{\mathcal{T}}$. 
\label{table:weights}
}
\end{table*}

\section*{\normalsize Mesoscale targeted interventions for epidemic spreading}
\label{sec:interventions}

\subsection*{\small Intervention design and evaluation}
The decomposition of a temporal network into a superposition of components, interpreted as mesostructures,
makes it possible to pinpoint and assess the contribution of individual mesoscopic features
to the overall structural and functional properties of the temporal network.
Indeed, the additive decomposition of Eq.~\ref{eq:decomp} allows to selectively remove
a chosen component $s$ by excluding the corresponding term,
yielding an altered temporal network,
\begin{equation}
\tilde{\mathcal{T}}^{s} = \sum_{r \neq s} \, \mathcal{S}_r \  ,
\label{eq:decomp_removed}
\end{equation}
which can be compared to the original network to elucidate the specific role of component $s$.
When component $s$ can be interpreted in terms of a specific behavioral pattern
(e.g., the lunch breaks of Fig.~\ref{fig:component}) its removal can be regarded
as the effect of an intervention strategy that selectively targets that behavior
(e.g., removing lunch breaks from the school schedule).

As a first step in exploring mesoscale interventions,
here we focus on the case study of the previous section
and investigate the effect of removing individual mesostructures
on the dynamics of simple epidemic processes unfolding over the school temporal network.
We study how the timing and size of the epidemic are affected
by the removal of individual network components, that is, by different targeted interventions
aimed at removing them.

We start with the simple case of a susceptible-infected (SI) process.
Each network node can be in either of two states: susceptible (S) or infected (I).
A susceptible node in contact with an infected one becomes infected
with a fixed probability $\lambda$ per unit time. Once infected, a node stays indefinitely in that state.
The system is initialized with all nodes in the S state, except for a single infected node (seed node).
The timing and duration of contacts between nodes is described by the temporal network $\tilde{\mathcal{T}}$,
which is a sequence of weighted networks: for each network snapshot, the weight of an edge between two nodes
gives the total duration of the contacts between those nodes during the corresponding time interval.
Although the SI process does not describe any realistic infectious disease,
it represents  a paradigmatic dynamical process
and it is frequently used to investigate the structural properties of temporal networks.

To investigate the effect on epidemic spread of removing a chosen mesostructure $\mathcal{S}_r$,
we simulate the epidemic process on the altered network $\tilde{\mathcal{T}}^r$,
and compare its dynamics to that observed on the full temporal network $\tilde{\mathcal{T}}$.
We quantify the effect of removing component $r$ by measuring the epidemic delay ratio~\cite{starnini2013immunization}
\begin{equation}
\tau_r = \left \langle \frac{t_{j}^{r} - t_j}{t_j} \right \rangle \, ,
\label{eq:delay}
\end{equation}
where $t_{j}$ is the half-infection time on $\tilde{\mathcal{T}}$ when the seed is node $j$,
and  $t_{j}^{r}$ is the same quantity for the altered temporal network $\tilde{\mathcal{T}}^r$.
The half-infection time is measured from the first time the seed node infects another node,
and the average $\langle \cdot \rangle$  is computed over all possible seed nodes $j$
and different starting times for the SI process (see Supplementary Information for details).

We also study the impact of mesoscale features on the dynamics of a more realistic epidemic process,
the Susceptible-Infected-Recovered (SIR) model.
In this model, $S \to I$ transitions occur as in the SI case when a susceptible node is in contact
with an infected one, with probability $\lambda$ per unit time.
Infected nodes recover with a constant rate $\mu$,
and Recovered (R) nodes no longer take part in the epidemic process. 
To assess the effect of removing a given mesostructure $r$
we compare the size of the epidemics at the end of the process,
measured as the final number of nodes that are in states I or R.
We compute the ratio $\rho$ between
the average size $\Omega^r$ of the epidemic on the altered network $\tilde{\mathcal{T}}^r$
and the average size $\Omega$ of the epidemic on the full network $\tilde{\mathcal{T}}$:
\begin{equation}
\rho_{\lambda,\mu}(r) = \frac{\langle \Omega^r \rangle}{\langle \Omega \rangle} \, .
\label{eq:ratio}
\end{equation}
The average is computed over all possible seed nodes and over different seeding times
near the beginning of the data set (see SI for details). 
Note that, due to the finite temporal span of our network data,
the epidemic process may not be over by the end of the simulation
(i.e., some nodes may still be in the I state).
In this case Eq.~\ref{eq:ratio} only yields an estimate
of how much the epidemic has been mitigated during the time interval covered by the data.

\subsection*{\small Targeted interventions: case study}
\begin{figure}[!htbp]
\centering
\includegraphics[width=\columnwidth, keepaspectratio]{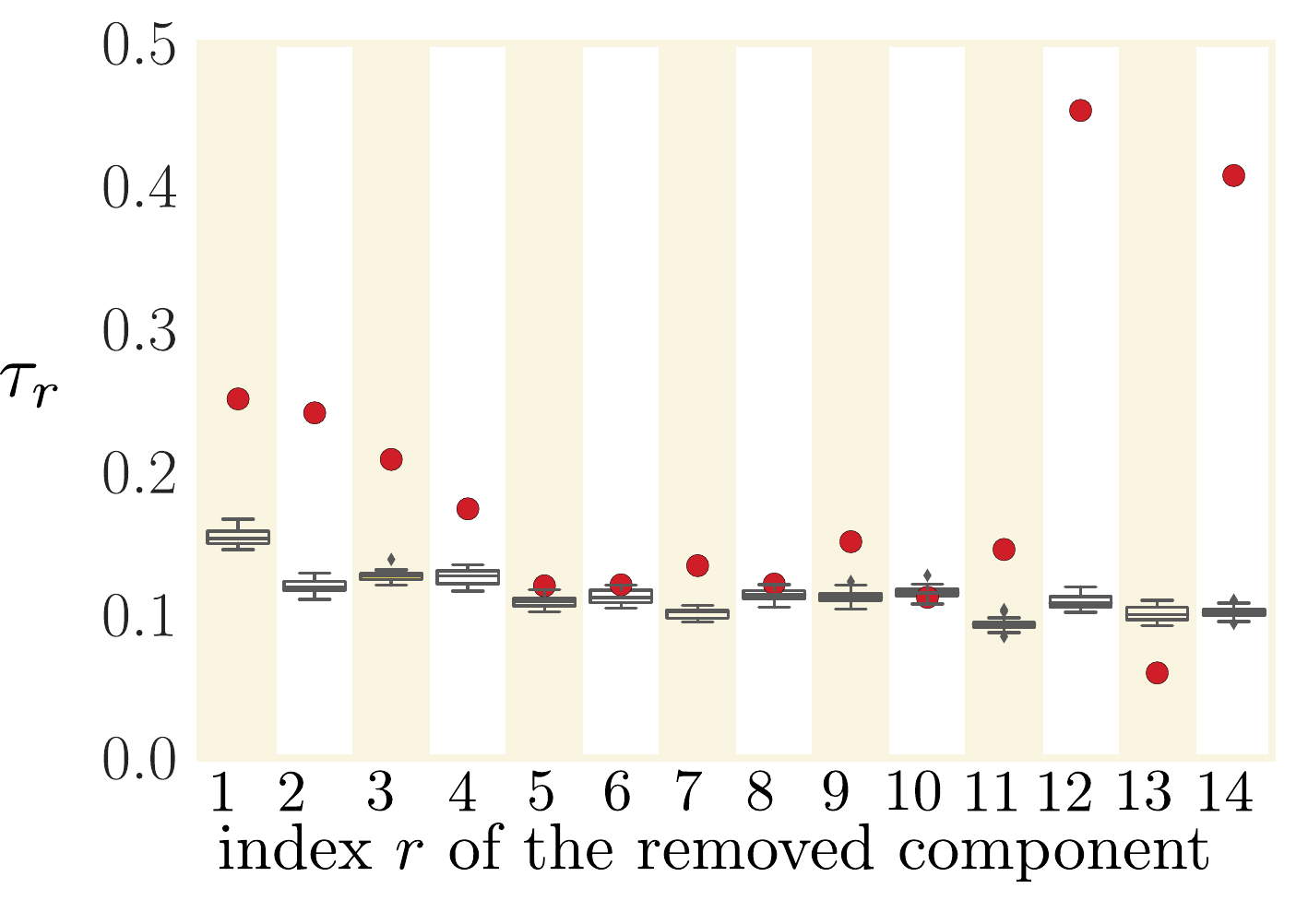}
\caption{
\textbf{Epidemic delay ratio as a function of removed component.}
The epidemic delay ratio $\tau_r$ of \protect{Eq.~\ref{eq:delay}} is plotted (solid circles)
as a function of the removed network component
and compared to the null model described in the main text (boxplots).
In each boxplot, the central band is the median,
the bottom and top of the box are the first and third quartiles.
Whiskers extent to the most extreme data point within 1.5 IQR of the inner quartiles.
Points beyond whiskers are outliers.
\label{fig:delay_rec_rem}
}
\end{figure}

We start from the decomposition of the school temporal network described in Fig.~\ref{fig:component}
and study the effect of removing individual mesostructures on the dynamics of an SI process.
We consider several values of the transmission rate $\lambda$:
here we present results for $\lambda=0.7$ and in the Supplementary Information
we show that qualitatively similar results hold for other values of the parameter.
Figure~\ref{fig:delay_rec_rem} shows the delay ratio $\tau_r$ of Eq.~\ref{eq:delay}
obtained by selectively removing each of the components $\mathcal{S}_r$, with $r = 1,\dots\ ,14$:
that is, comparing the dynamics of the SI process over each of the altered networks $\tilde{\mathcal{T}}^r$
with the dynamics observed for the full temporal network $\tilde{\mathcal{T}}$.
Since the altered networks have by definition a smaller total weight $\Vert \tilde{\mathcal{T}}^r \Vert$
than the full network, we expect epidemic spread to be mitigated for the altered networks.
To assess whether a given component $r$ plays a structural role for epidemic spread
that goes beyond its weight, it is thus important to compare the delay ratio $\tau_r$ observed
on removing that component with a suitable null model. Such a null model is obtained
by measuring the delay ratio for several stochastic realizations of altered temporal networks
obtained from $\tilde{\mathcal{T}}$ by removing, at random, a fraction of weight
equal to that of component $r$ (as given by Table~\ref{table:weights}).
The distributions of $\tau_r$ for the null models are shown by the boxplots of Fig.~\ref{fig:delay_rec_rem}.

We observe that the delay ratio obtained by selectively removing a component (solid circles)
is almost always larger than the one obtained by removing links at random (boxplots):
Such targeted interventions have a stronger effect than random removals of links.
For the first $9$ structures, the delay ratio is positively correlated with the component weight of Table~\ref{table:weights},
as we could naively expect.
Strikingly, the removal of components $12$ or $14$ slows down the epidemic spread considerably.
These two mesostructures, despite carrying a smaller total weight than most other components (see Table \ref{table:weights}),
thus appear to play a more important role for SI spreading. As shown in Fig.~\ref{fig:component}
both components $12$ and $14$  mix nodes from different classes and have activity concentrated
during the lunch break of the first day. The other components correspond either to individual classes
 (components $1$ through $9$ and $11$) or to components, such as $10$ and $13$, with little activity on the first day.
Overall, the SI process is slowed down the most by removing comparatively weak structures (in terms of weight)
that mix classes and occur early in the data.
We remark that the most effective intervention for mitigating epidemic spread
thus involves the removal of seemingly minor mesostructures that correspond
to complex correlations between link activity and network structure,
and cannot be uncovered by standard community detection approaches.

%
We now turn to the case of an SIR process
and investigate the effect of mesostructure removal
on the dynamics of an SIR epidemic over the temporal network.
We compute the epidemic size ratio $\rho_{\lambda,\mu}(r)$ of Eq.~\ref{eq:ratio}
for a wide range of values of the parameters $\lambda$ and $\mu$, and for all components $r$.
The results are shown in Fig~\ref{fig:heatmap} for selected components
and in the Supplementary Information for all the other components.
%
%
%
\begin{figure*}[!ht]
\centering
\includegraphics[width=\textwidth, keepaspectratio]{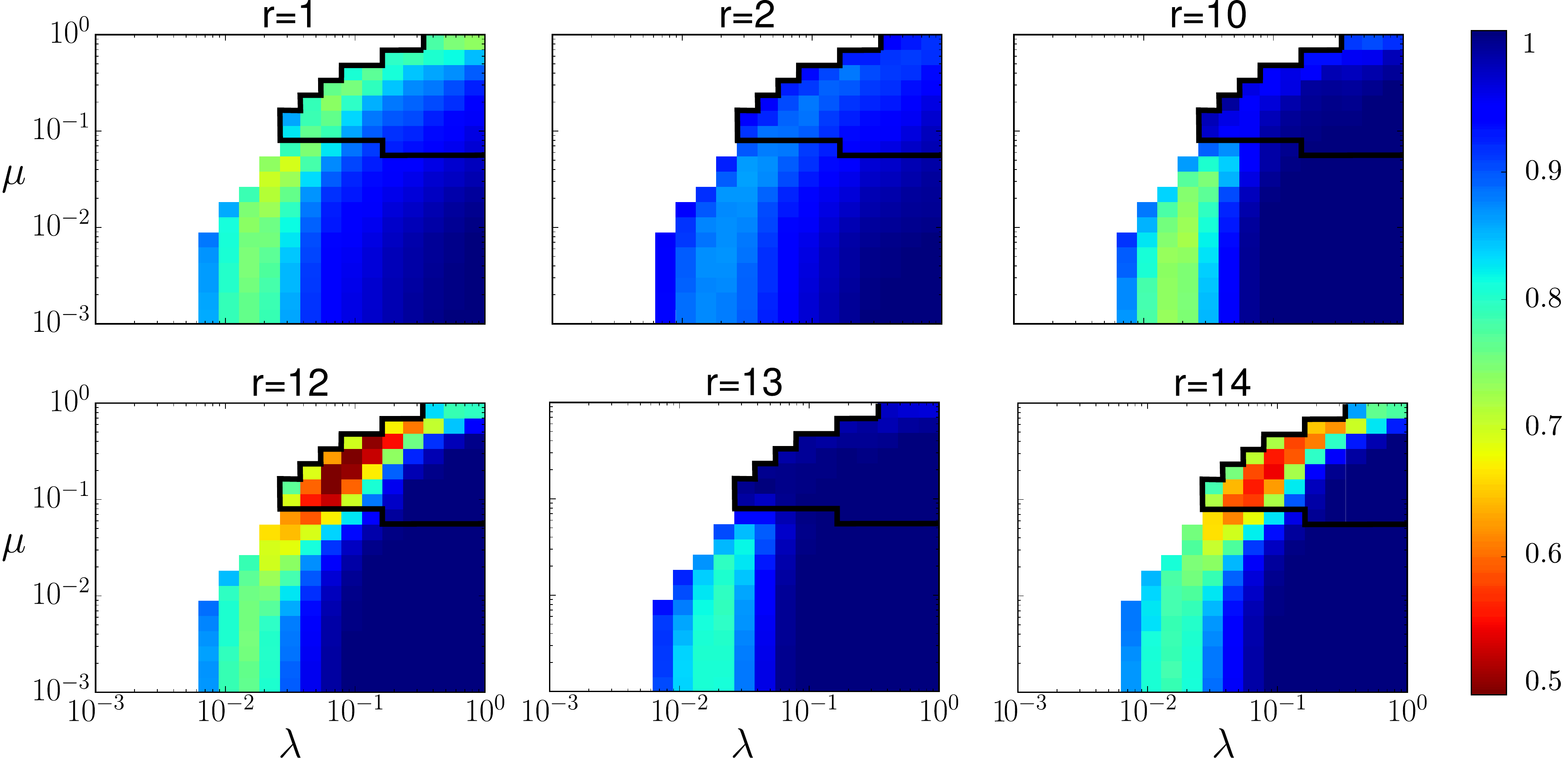}
\caption{
\textbf{Epidemic size ratio as a function of SIR model parameters.}
Each heat map corresponds to a targeted intervention that selectively removes one component.
For each removed component $r$ the heat map shows the epidemic size ratio $\rho_{\lambda,\mu}(r)$
as a function of the SIR parameters $\lambda$ and $\mu$.
$\rho_{\lambda,\mu}(r) = 1$ indicates that the intervention does not affect epidemic size.
The white area is the region where the epidemic dies out,
i.e., it fails to affect more than 1\% of the network nodes.
The region inside the black contour line corresponds to parameter values
such that the SIR epidemic finishes within the finite span of the school dataset (2 days),
both for the full and the altered temporal networks. That is, for those parameter values
the epidemic size ratio is not affected by the finite temporal span of the dataset.
\label{fig:heatmap}
}
\end{figure*}
We notice that for large $\lambda$ values and small $\mu$ values (bottom right of the heat maps)
the epidemic size ratio $\rho_{\lambda,\mu}(r) \simeq 1$:
in this region the epidemics spreads and finishes fast, hence mesostructure removal cannot mitigate its size.
The parameter region where mesostructure removal can have a strong effect,
therefore, is the arc-shaped region visible, e.g., in the heat map for $r=1$ (lighter blue/green).
The removal of class components $r=1$ and $r=2$ mildly mitigates the epidemic
for a broad range of parameter within that region (almost no effect is obtained for the other class components,
as shown in Fig.~S4 of the Supplementary Information).
Removing the class component $1$ has a comparatively higher effect because
of the higher overall weight of that component (Table~\ref{table:weights}).

\begin{figure}[!htbp]
\centering
\includegraphics[width=0.8\columnwidth, keepaspectratio]{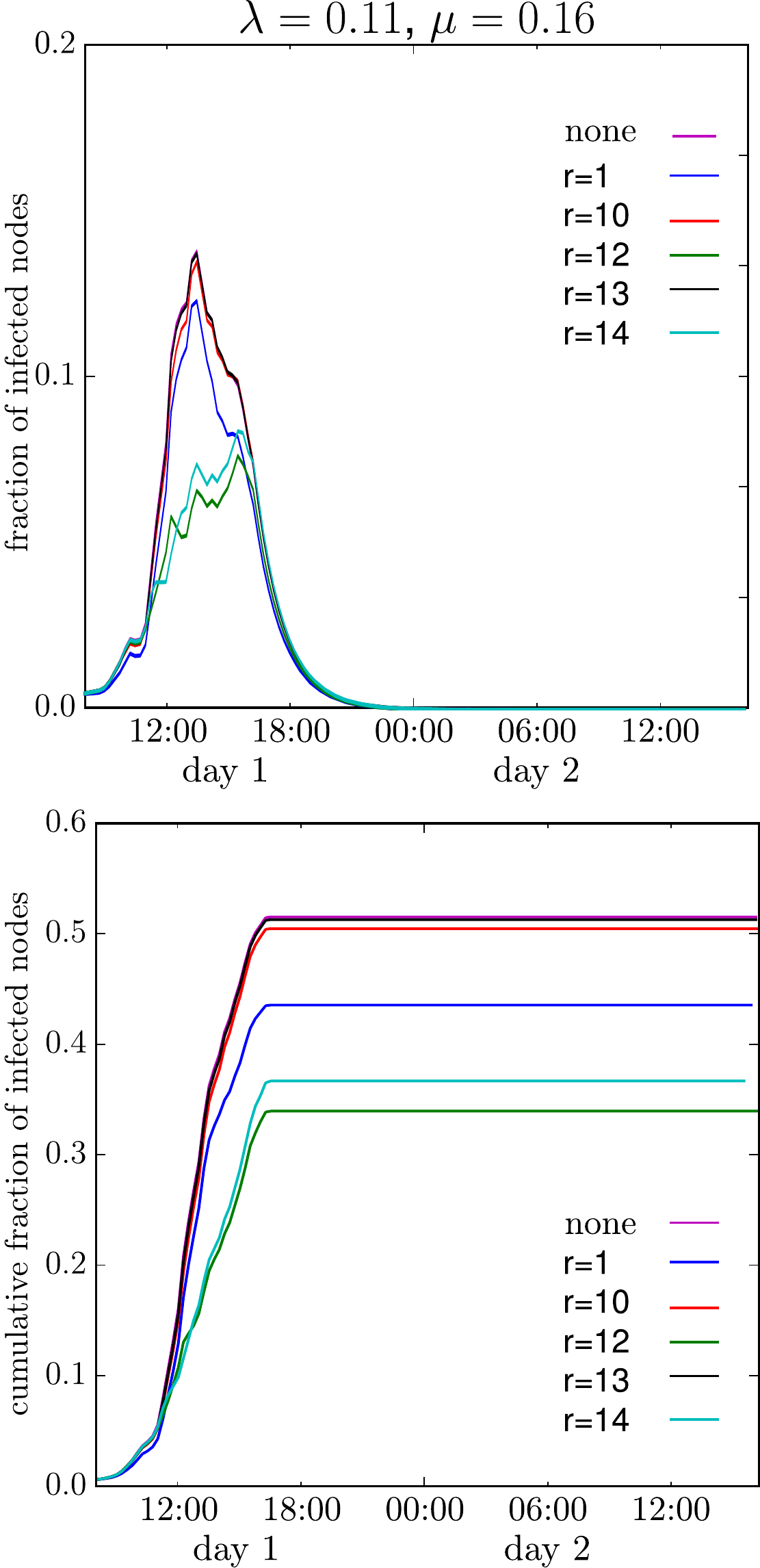}
\caption{
\textbf{Fraction of infected nodes as a function of time for a sample parameter choice.}
Temporal evolution of the average fraction of infected nodes (top)
and of the cumulative fraction of infected nodes (bottom) as a function of time
for a sample parameter choice of $\lambda=0.11$ and $\mu=0.16$.
The curve marked as \textit{none} corresponds to the spread on the unmodified temporal network $\tilde{\mathcal{T}}$.
The other curves show the effect of selectively removing components $r=1,10,11,12,14$.
For the chosen parameter values, epidemic spread occurs on the first day,
hence removing the early class-mixing components $r=12$ or $r=14$ has a strong mitigation effect.
\label{fig:reg_evolution}
}
\end{figure}

On taking a closer look at Fig.~\ref{fig:heatmap} we notice that, depending on the removed component,
significant epidemic mitigation can be achieved in two main parameter regions within the arc.
The first region falls within the black contour (high values of $\lambda$ and $\mu$)
and corresponds to parameter values yielding epidemics that finish by the end of the second day.
In this region the epidemic can be strongly mitigated ($\rho_{\lambda,\mu} \simeq 0.5$)
by removing mesostructures that involve mixing of classes on the first school day, namely $r=12$ and $r=14$.
This is consistent with the results of Fig.~\ref{fig:delay_rec_rem} for the SI case.
Conversely, removing the class-mixing components $r=10$ and $r=13$,
which are mostly active on the second day (Fig.~\ref{fig:component})
has a negligible effect in the same parameter region.
For a sample parameter choice within this region, in Fig.~\ref{fig:reg_evolution}
we illustrate the temporal evolution of the epidemics and the effect of removing individual components:
the epidemic on the unmodified network peaks on the first day and, consistently with the above remarks,
removing $r=12$ or $r=14$ strongly mitigates the fraction of affected nodes,
whereas the removal of $r=10$ and $r=13$ has a negligible effect.
We remark that removing the early class-mixing mesostructures also mitigates the SIR epidemic
for a wide range of parameter values outside the black contour, as visible in Fig.~\ref{fig:heatmap},
and that removing these mesostructures is more effective, in general, than removing
class components, even though the latter account for a larger network weight.
The second parameter region of interest lies in the lower part of the arc (low values of $\lambda$ and $\mu$)
and is visible in Fig.~\ref{fig:heatmap} for $r=10$ and $r=13$.
This region corresponds to parameter values for which the SIR process is slower
and the epidemic does not finish over the 2-day temporal span of the data.
In this case, removing either of the class-mixing mesostructures  $r=10,12,13,14$
as a similar and limited effect on the epidemic size reached at the end of the second day.

\section*{\normalsize Discussion}
We have put forward a methodology to systematically assess the relevance of mesostructures
for dynamical processes such as epidemic spread in temporal networks.
The proposed method only uses time-resolved connectivity information without metadata
and allows to uncover complex patterns of correlated link activity
whose removal has a strong impact on simulated epidemic spreading for a broad range of model parameters.
The removal of an individual mesostructure can be regarded as a targeted  intervention at the mesoscale,
such as the removal of a specific group of nodes or the suppression of a specific activity pattern
(e.g., a lunch break). These targeted interventions have a potentially high relevance,
as they do not involve the system as a whole, nor they need actions at microscopic level of individual nodes.
Hence, they might strike a balance between cost and effectiveness
that might make them actionable in a variety of applications.

As an illustration of the method, we have considered a case study where simple but paradigmatic
spreading processes occur on top of an empirical temporal network describing face-to-face interactions in a school.
In this case, the most important mesostructures for epidemic spread
are not those involving most of the link activity, but instead consist of weaker, temporally-localized mixing patterns
corresponding to scheduled social activities.
It is important to remark that these mesostructures cannot by found by means of community detection techniques
on static networks, and that our methods reveals them in an unsupervised fashion.
We also notice that the actionability of an intervention that suppresses a given mesostructure
hinges on the recognizability of that mesostructure in terms of relevant contextual information,
such as a known or understandable grouping of nodes, 
temporal activity localized at specific times of a known schedule,
or a specific relation to other metadata for the context at hand.
For example, the selective removal of component $\mathcal{S}_{12}$ of Fig.~\ref{fig:component}
only becomes an actionable targeted intervention when its temporal activity profile is
identified as a mixing pattern associated with the lunch break on the first school day.

Our work is a first step towards the systematic design of mesoscale interventions in temporal networks.
Natural future steps include the use of our methodology on different temporal networks,
both empirical and synthetic, the investigation of the impact of incomplete or noisy data, 
and the study of other dynamical processes.


\section*{\normalsize Methods}
A 3-mode tensor $\mathcal{T}$, with entries $T_{ijk}$, can be approximated by a sum $\tilde{\mathcal{T}}$
of $R$ rank-$1$ tensors~\cite{Cichocki,Kolda2009,Morup11},
\begin{equation}
\tilde{\mathcal{T}} = \sum_{r=1}^R \, \mathcal{S}_r \ = \sum_{r=1}^R \, \mathbf{a_r} \circ \mathbf{b_r} \circ \mathbf{c_r} \  ,
\label{eq:approx}
\end{equation}
where each component tensor $\mathcal{S}_r$ is the outer products of three vectors,
$\mathcal{S}_r = \mathbf{a_r} \circ \mathbf{b_r} \circ \mathbf{c_r}$.
The Frobenius norm of the difference between $\mathcal{T}$ and $\tilde{\mathcal{T}}$,
$\epsilon = \Vert \mathcal{T} - \tilde{\mathcal{T}} \Vert$,
quantifies the error in approximating the original tensor with the sum $\tilde{\mathcal{T}}$.
The number of components $R$ is interpreted as the number of sought mesostructures
in the decomposition of the temporal network. Its value needs to be set by balancing
the conflicting goals of minimizing $\epsilon$, i.e., recovering as much as possible
of the original tensor, and avoiding overfitting~\cite{Bro,Kolda2009,gauvin2014detecting}. 
The sets of vectors
$a_{\{1,2,\dots,R\}}$, $b_{\{1, 2, \dots, R \}}$, $c_{\{1, 2, \dots, R \}}$
can be combined into matrices $\mathbf{A} \in \mathbb{R}^{N \times R}$,
$\mathbf{B} \in \mathbb{R}^{N \times R}$ and $\mathbf{C} \in \mathbb{R}^{S \times R}$.
These matrices all have $R$ columns,
one for each component of the decomposition, i.e., one for each sought mesoscale structure.

For an undirected temporal network (such as the one of our case study)
the adjacency matrix represented on each tensor slice is symmetric
and the decomposition yields $\mathbf{A} \simeq \mathbf{B}$.
In practice it is possible to directly seek a decomposition with $\mathbf{A} \equiv \mathbf{B}$,
i.e., with component tensors of the form $\mathcal{S}_r = \mathbf{a_r} \circ \mathbf{a_r} \circ \mathbf{c_r}$.
Hence in the main text we only refer to the matrices $\mathbf{A}$ and $\mathbf{C}$:
the matrix elements $a_{ir}$ of $\mathbf{A}$ associate each component $r = 1,\dots,R$
with individual nodes of the network,
while the matrix elements $c_{kr}$ of $\mathbf{C}$
encode the temporal activity pattern of each component, with time indexed by $k$.
The decomposition of Eq.~\ref{eq:approx} can be written as
\begin{equation}
T_{ijk} = \sum_{r=1}^R \, a_{ir} \, a_{jr} \, c_{kr} \  .
\label{eq:approx2}
\end{equation}
The matrix elements $a_{ir}$ indicate notion of membership
of node $i$ to component $r$, so this representation is suitable to describe
overlapping mesostructures and complex correlations between connectivity patterns
and activity patterns over time. Notice that even when the original temporal network
is unweighted, that is, the tensor $\mathcal{T}$ is binary-valued,
the component tensors $\mathcal{S}_r$ correspond in general to weighted networks,
and the approximating tensor $\tilde{\mathcal{T}}$ is in general real-valued.

To compute the decomposition of Eq.~\ref{eq:approx}
we need to solve an optimization problem that minimizes the residual $\epsilon$.
A standard way to do so is to convert the $3$-mode problem yielded by Eq.~\ref{eq:approx}
into three coupled $2$-mode sub-problems: This is done by unfolding the original tensor $\mathcal{T}$
along its three modes, a technique also known as matricization~\cite{Cichocki}.
Using this representation, the original tensor approximation problem is cast
into $3$ coupled matrix approximation problems
that involve the factor matrices $\mathbf{A}$, $\mathbf{B}$, and $\mathbf{C}$.
Since the corresponding optimization problems are convex with respect to either of the matrices
-- but not with respect to them all -- it is possible to solve the optimization
through a method known as Alternative Non-negative Least Squares~\cite{Paatero}.
The minimization of $\epsilon$ is usually carried out with non-negativity and/or sparsity constraints
on the factor matrices~\cite{Cichocki,Kolda2009}, as this is known to yield decompositions
that can be interpreted as parts-based representations of the original data~\cite{lee1999learning}.

Here we carry out the decomposition with non-negativity constraints,
using Alternating Non-negative Least Squares and a block-coordinate descent method
to improve convergence speed~\cite{Kim2012,kim2014algorithms}.
Our implementation builds on the MATLAB code of ref.~\citenum{ParkNTFcode}.
The number of components $R$ is set by using the core consistency diagnostic~\cite{Bro,gauvin2014detecting} 
optimized over $5$ stochastic realizations of the decomposition
that differ for the random initial conditions of the factor matrices.
The components $\mathcal{S}_r$ are ordered so that the Euclidean norm
of $\mathcal{S}_r$ decreases for increasing $r$.  That is, the first column
of matrices $\mathbf{A}$, $\mathbf{B}$ and $\mathbf{C}$ correspond to the component
with highest Euclidean norm, and successive columns encode weaker and weaker components.

\section*{\normalsize Acknowledgements}
The authors acknowledge support from the Lagrange Project of the ISI Foundation funded by the CRT Foundation
from the Q-ARACNE project funded by the Fondazione Compagnia di San Paolo,
and from the FET Multiplex Project (EU-FET-317532) funded by the European Commission.
The authors acknowledge help from Marco Quaggiotto for the design of figure panels.

\section*{\normalsize Author Contributions}
All authors contributed to the manuscript. 
LG, AB and CC designed the study. AB and CC supervised the study.
AB and CC collected and post-processed the data.
LG and AP analyzed data and prepared figures.
LG carried out computer simulations.
LG, AB, CC wrote the manuscript. 
All authors reviewed the manuscript.

\section*{\normalsize Additional Information}
The authors declare no competing financial interests.

\bibliographystyle{naturemag}

\clearpage

\renewcommand{\thefigure}{S\arabic{figure}} 
\section*{Supplementary Information}
\setcounter{figure}{0}   
\section*{\normalsize Properties of the decomposition of the contact network}
\begin{figure}[!htbp]
\centering
\includegraphics[width=0.8\linewidth]{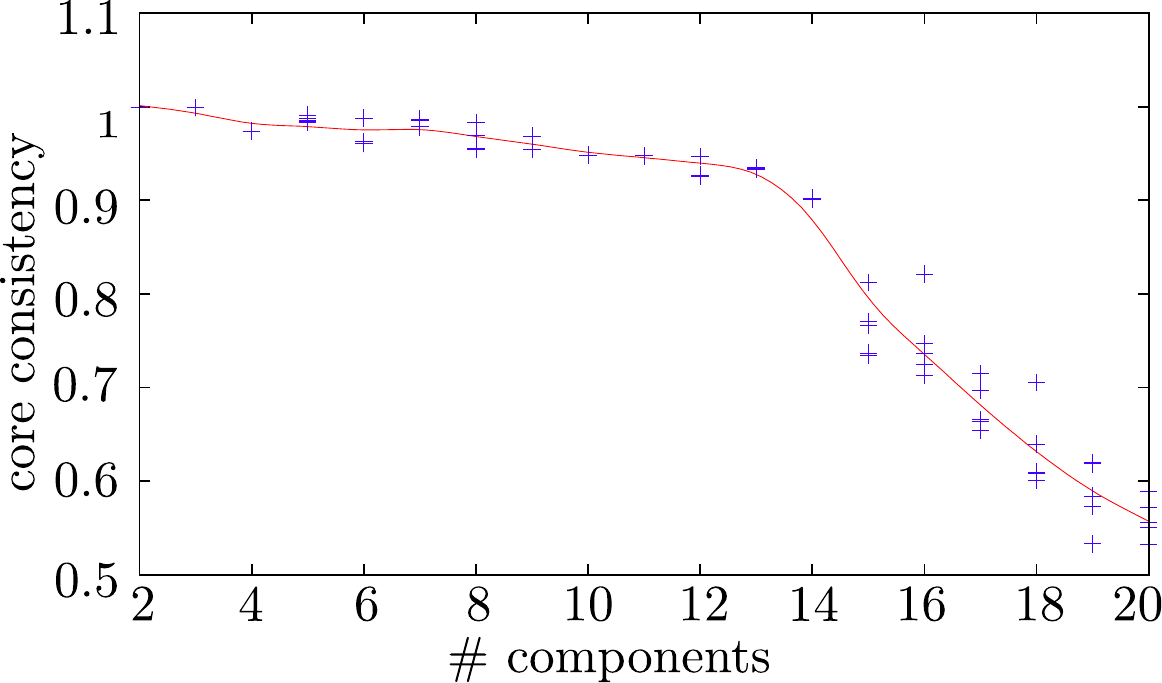}
\caption{\textbf{Core consistency curve with respect to the number of components}. Each cross corresponds to a decomposition, $5$ realizations are shown
for each number of components used. The line acts as a guide for the eyes.}
\label{core_consistency}
\end{figure}

\begin{figure}[!htbp]
\centering
\includegraphics[width=0.8\linewidth]{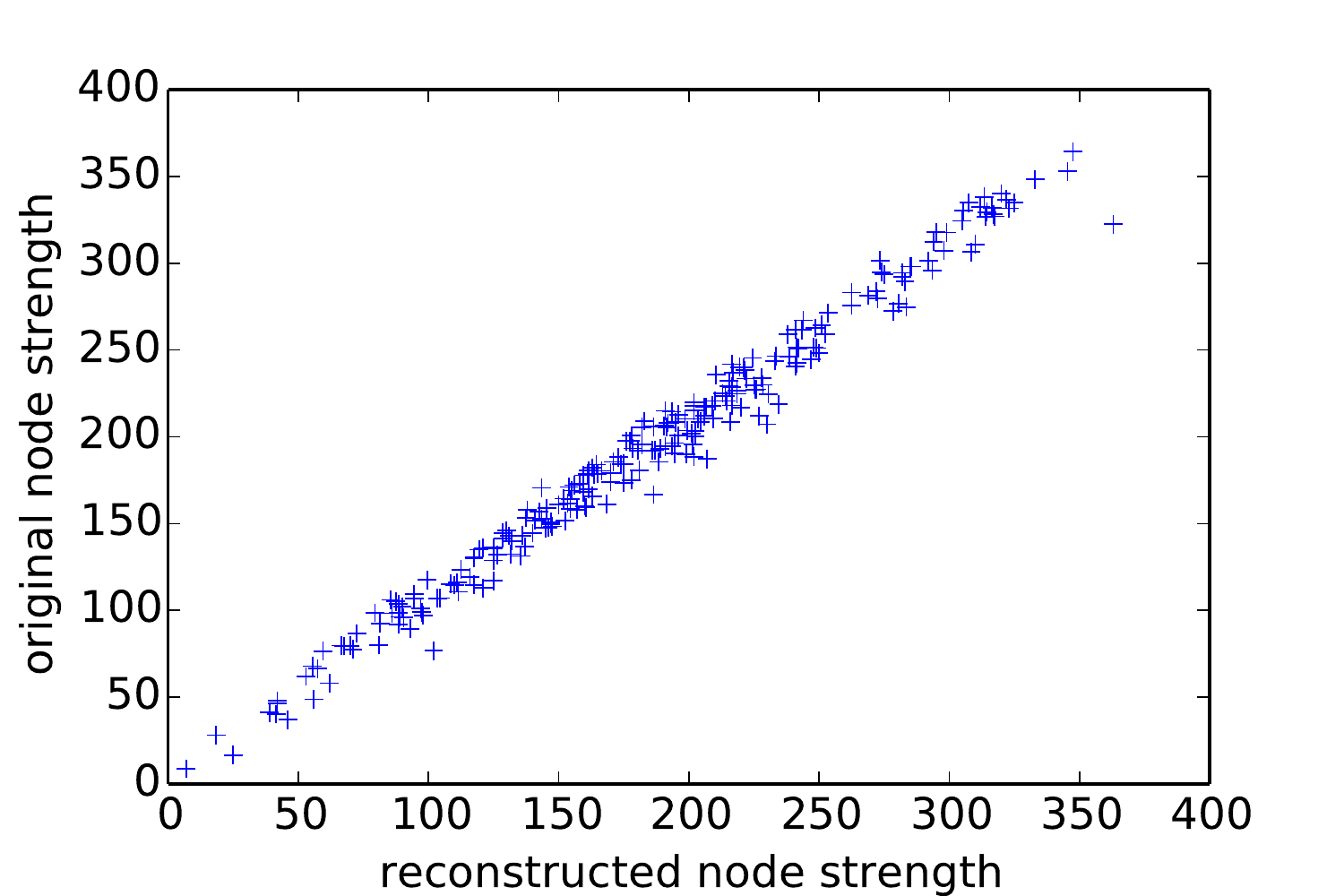}
\caption{\textbf{Cumulative node strength} in the original temporal network described by $\mathcal{T}$, versus cumulative node strength
in the reconstructed tensor $\tilde{\mathcal{T}}$ with $R=14$ components.}
\label{act_nodes_rec_or}
\end{figure}

The original temporal network $\mathcal{T}$
is decomposed by the non-negative tensor factorization method described in the Methods section and 
approximated by the resulting tensor $\tilde{\mathcal{T}}$, which is a sum  of $R$ products of 
lower-dimensional factors.
This approximation was carried out for several values of the number $R$ of components, i.e. of rank-$1$ tensors,
in order to determine the range of values such that $R$ is large enough to describe the mesoscale properties of the original network 
and small enough to avoid overfitting. The core consistency, a measure introduced in \cite{Bro}  and 
plotted in Fig. \ref{core_consistency} for several values of the number $R$ of rank-$1$ tensors, guides the choice 
of a representative number of components for $\tilde{\mathcal{T}}$.
In the case study we consider, we limit the decomposition to $R=14$ terms  to avoid overfitting. 
In Fig. \ref{act_nodes_rec_or}, we show that this number of terms is sufficient  
by comparing the total activity (number of interactions in the whole data set) of
each node in the approximated network $\tilde{\mathcal{T}}$ and in the original network $\mathcal{T}$.
As mentioned in the Methods section, $\tilde{\mathcal{T}}$ is real-valued even if $\mathcal{T}$ is binary-valued.

%
\section*{\normalsize Spreading processes and intervention strategies}
Both Susceptible-Infectible and Susceptible-Infectible-Recovered processes are run on the reconstructed network and the altered networks. 
We consider $R=14$ altered networks: the  $r-$th  one, $\tilde{\mathcal{T}}^r$, is built by excluding the  $r-$th component from the sum 
defining  $\tilde{\mathcal{T}}$.


\subsection*{\small Details of the epidemics simulations}
The quantities measured for the SI process are averaged, for each altered network and for each parameter $\lambda$, over
all possible seeds and for each seed over $50$ spread starting times taken 
between $10\%$ and $30\%$ of the total time length of the dataset. 
The quantities measured in the case of the SIR process are averaged, for each altered network and 
each set of parameters $(\lambda,\mu)$, over all possible seeds and for each seed 
over $10$ starting times taken between 
$10\%$ and $15\%$ of the total time length of the dataset.

\subsection*{\small Efficiency of targeted interventions for the SI process: effect of different propagation probabilities}

Figure 2 of the main text shows the delay ratio of SI processes obtained by the selected removal of each mesostructure of the temporal network,
for a specific value of the spreading rate $\lambda$. The maximal value of $\lambda$ tested is chosen such that 
 $\lambda$ times the maximal weight of the links (here $1.1$) does not exceed $1$. 

\begin{figure*}
    \centering
    \subfigure[]{\includegraphics[width=0.48\textwidth]{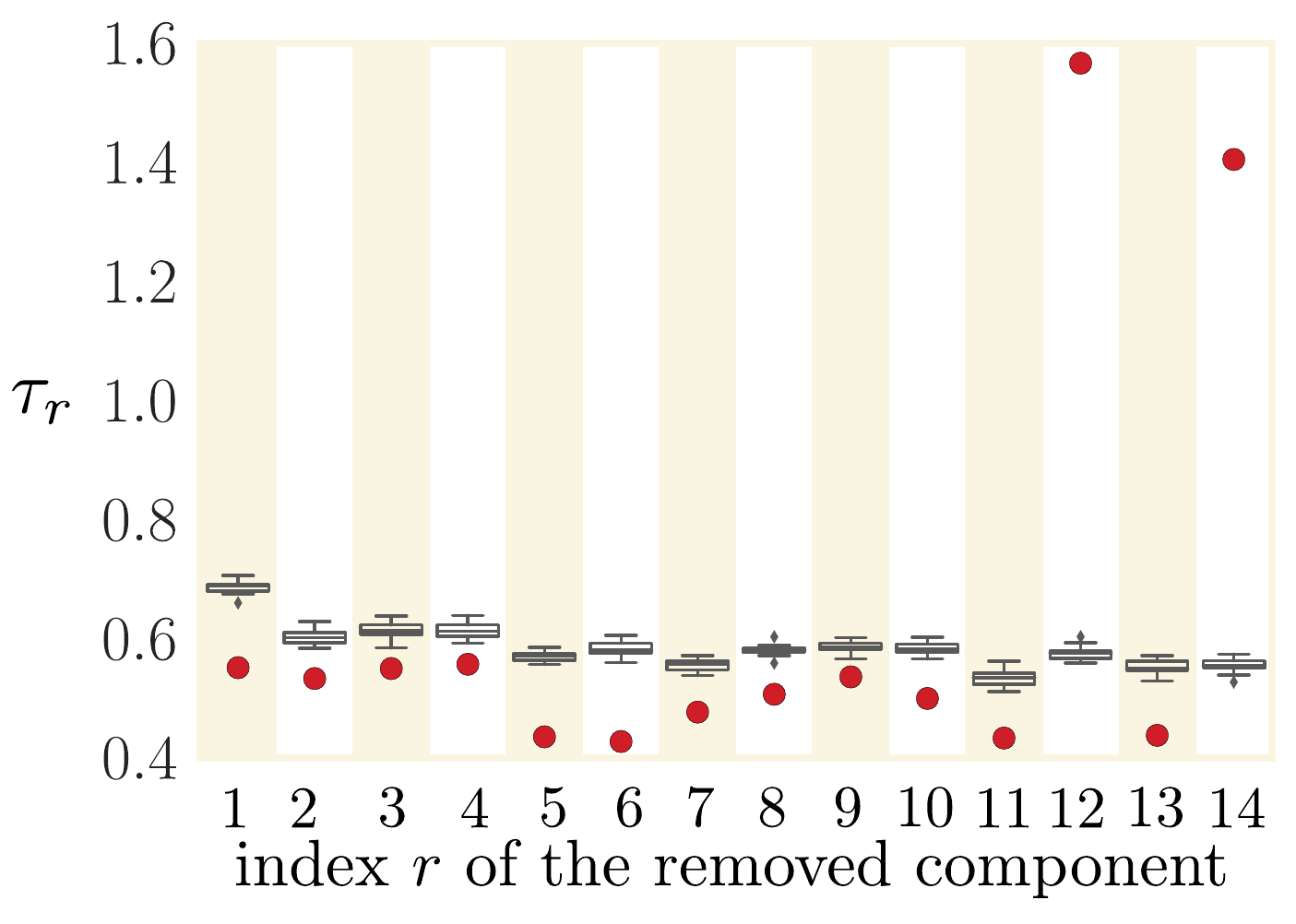}}
    \subfigure[]{\includegraphics[width=0.48\textwidth]{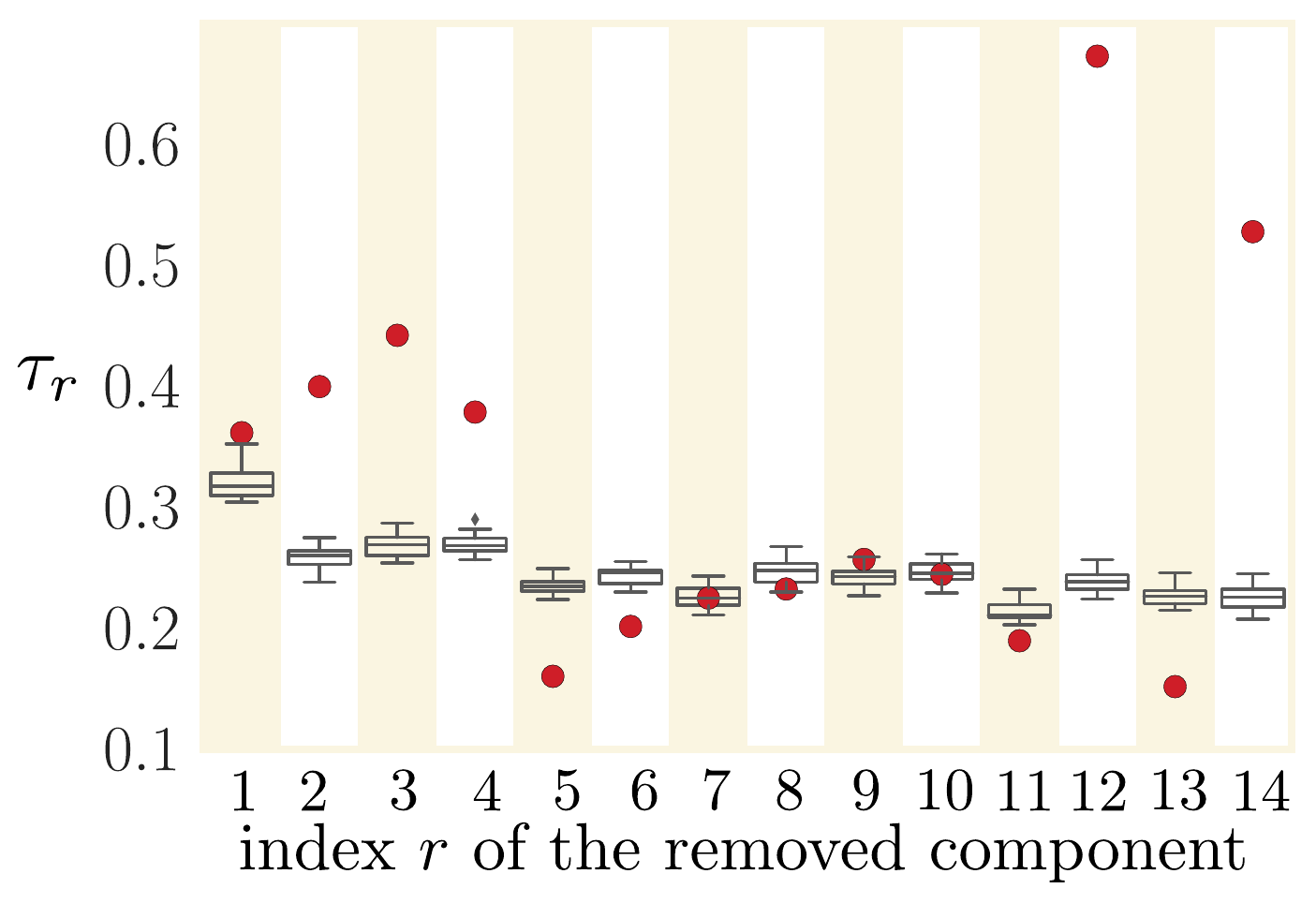}}
    \subfigure[]{\includegraphics[width=0.48\textwidth]{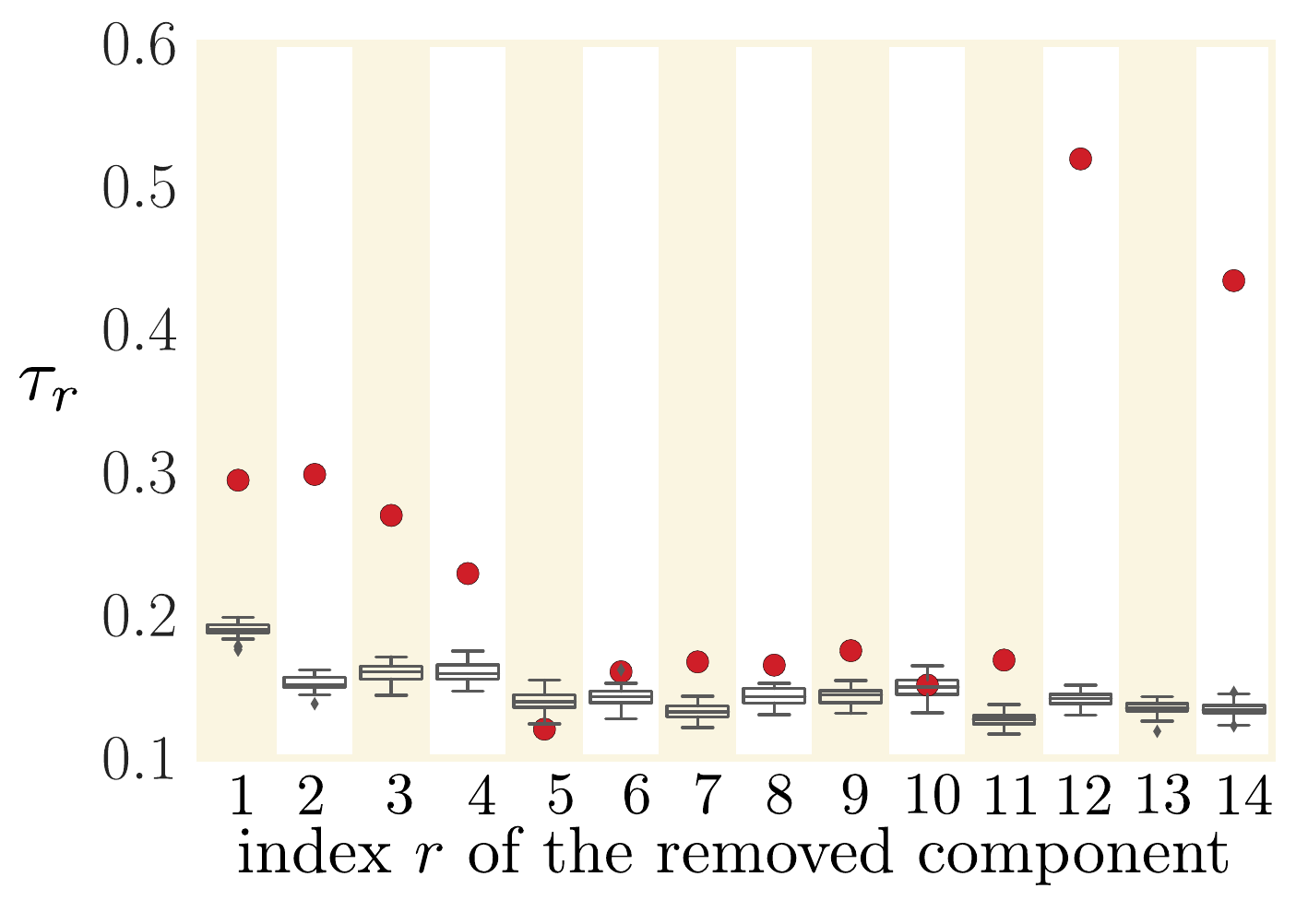}}
    \subfigure[]{\includegraphics[width=0.48\textwidth]{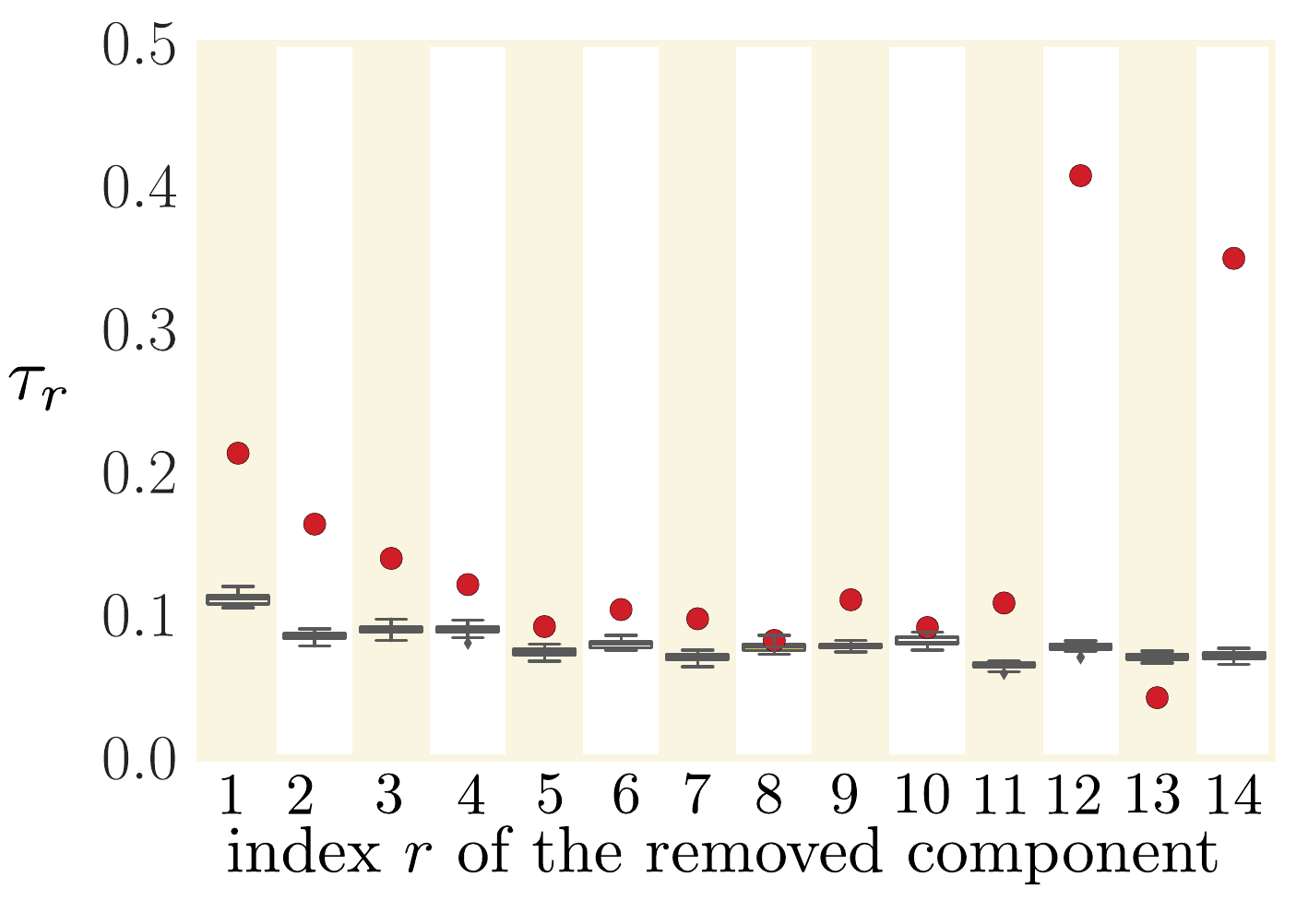}}
   \caption{\textbf{Measured delay ratios for SI processes simulated on top of the reconstructed network 
   $\tilde{\mathcal{T}}$ and of the $14$ altered networks.} The boxplots correspond to the delay ratios obtained
by simulating SI processes on top of $\tilde{\mathcal{T}}$ and of 
$14$ ensemble of networks altered by random removal of interactions: for each mesostructure $r$, we 
remove from  $\tilde{\mathcal{T}}$ weights at random under the constraint that  
the sum of the  removed weights is equal to the sum of the weights of the component $r$. Each boxplot gives the 
distribution of delay ratios obtained for $20$ realizations of the random removal procedure.
Horizontal lines inside each box show the median values,
and each box extends from the first to the third quartile of the corresponding distribution. 
The reach of whiskers are determined by the sum of the third quartile and 
the interquartile weighted by 1.5. Points beyond the whiskers are outliers. 
The transmission probability of the SI processes is  (a) $\lambda=0.1$, (b) $\lambda=0.3$, (c) $\lambda=0.5$, (d) $\lambda=0.9$.}
    \label{fig:delay_ratio}
\end{figure*}

We show in Fig. \ref{fig:delay_ratio} the results obtained with different values of $\lambda$ on each altered network together with 
the results obtained 
on $20$ realizations of associated null models. Each null model associated to an altered network is built by taking the fully reconstructed 
network $\tilde{\mathcal{T}}$ and
zeroing out link weights at random until the sum removed is equal to the actual cumulated weights removed in the corresponding altered 
network $\tilde{\mathcal{T}}^r$.
Although the efficiency of the single removal of each mesostructure depends on $\lambda$, the global picture remains unchanged:
the removal of one of the mixing patterns active during the first day ($r=12$ or $r=14$) largely outperforms the removal of any of the other
mesostructures as well as the random removal of an equivalent amount of activity. 

Let us also note for precision that  the half-infection time used in the computation of the infection delay ratio is measured starting from the time at which the
seed infects a node instead of the time of its first contact (as it was done in  Ref.~\citenum{starnini2013immunization}).
The reason is that the links of the reconstructed network are not binary (active or inactive) but carry weights, so that
the precise definition of a contact is blurred and recovering a binary definition of contacts would require to set an arbitrary threshold.

\subsection*{\small Targeted intervention for the SIR process}
To quantify the difference between SIR processes unfolding on top of $\tilde{\mathcal{T}}$ and on top of altered networks, 
we measured the averaged ratio, $\rho_{\lambda,\mu}(r)$,
between
the average size $\Omega^r$ of the epidemic on the altered network $\tilde{\mathcal{T}}^r$
and the average size $\Omega$ of the epidemic on the full network $\tilde{\mathcal{T}}$. 
To estimate the statistical significance  of our results, we have performed
 Wilcoxon signed-rank tests for each parameter set ($\lambda,\mu,r$) comparing the epidemics size 
 distributions pairwise between simulations performed on each altered network and on the full $\tilde{\mathcal{T}}$.
 The test's result shows that either the averages of the epidemic sizes are significantly different with a $p-$value lower
 than $10^{-3}$, or the ratio
 $\rho_{\lambda,\mu}(r)$  lies between $0.99$ and $1$. Therefore, this test shows that whenever
 $\rho_{\lambda,\mu}(r)$ is smaller than $0.99$, 
 the observed difference between the outcomes of the spreading processes run on $\tilde{\mathcal{T}}^r$ and $\tilde{\mathcal{T}}$
is significant. 

\begin{figure*}[!htbp]
 \centering
\includegraphics[width=\textwidth, keepaspectratio]{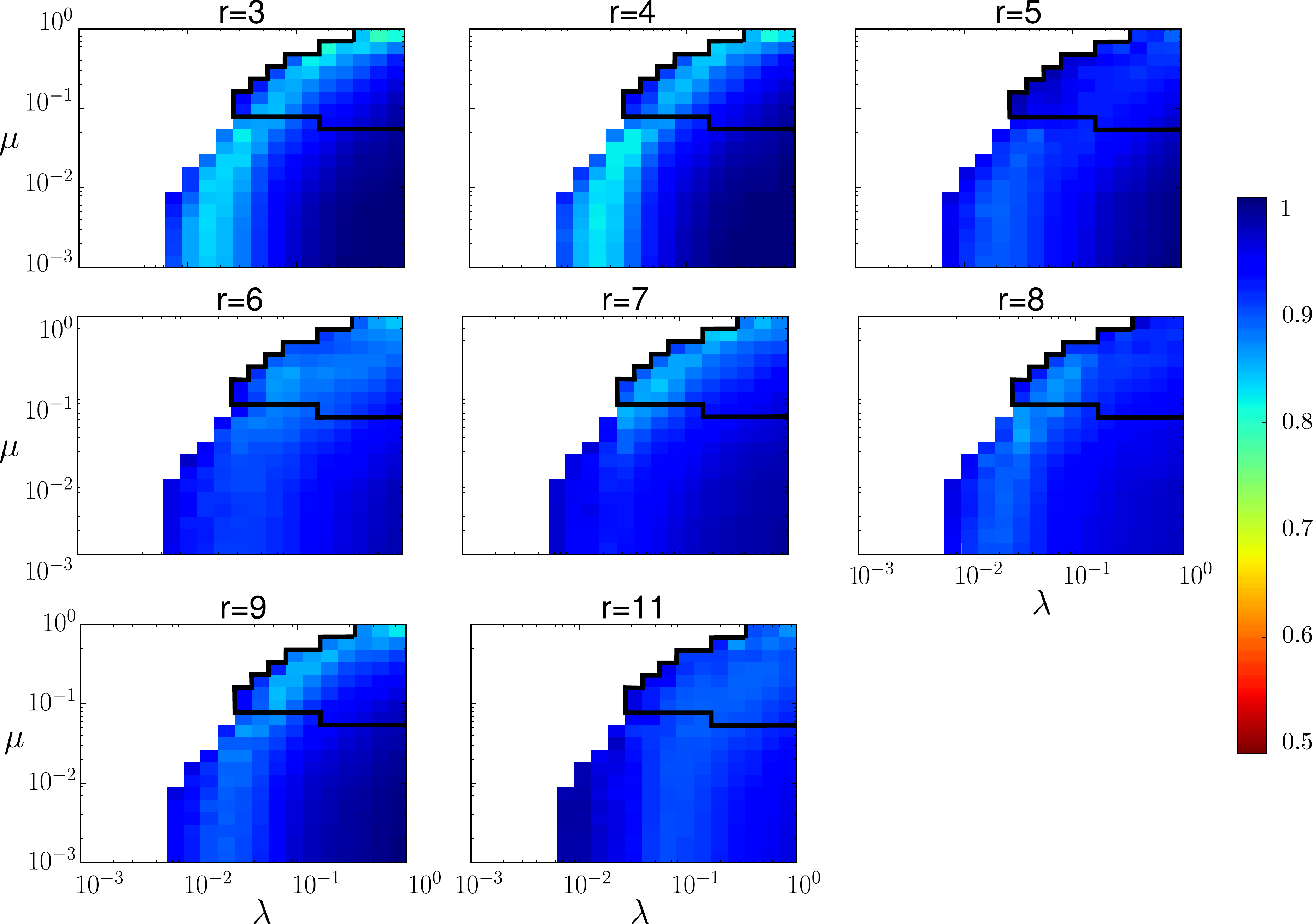}
\caption{\textbf{Ratio of epidemics sizes $\rho_{\lambda,\mu}(r)$  for SIR processes
simulated on top of altered networks with a mesostructure $r$ removed
and on top of the whole  $\tilde{\mathcal{T}}$}. The values of $r$  selected here are those which were not on Fig. 3 of the main text. 
The white part of each heat map corresponds to the region where the simulated epidemic process dies out. 
The black contour corresponds to the boundary of the parameter region such that the SIR process is over within the finite span of the dataset
(i.e., to parameter values yielding epidemic processes that last at most two days): in this parameter region, the 
finite length of the dataset has no influence on the outcome of the epidemic.
}
\label{fig:secondary_heatmap}
\end{figure*}

We show in Fig. \ref{fig:secondary_heatmap} 
the ratio of epidemic sizes $\rho_{\lambda,\mu}(r)$ for SIR processes
simulated on top of altered networks with a mesostructure $r$ removed
and on top of the whole decomposition $\tilde{\mathcal{T}}$, for the values of $r$ not shown in the main text. 
These mesostructures describe activities occurring in single classes, and their removal  have only negligible impact on the outcome of the 
 SIR process.

\subsection*{\small Evolution of the epidemic size in different regimes}

\begin{figure*}
    \centering
    \subfigure[]{\includegraphics[width=0.49\textwidth]{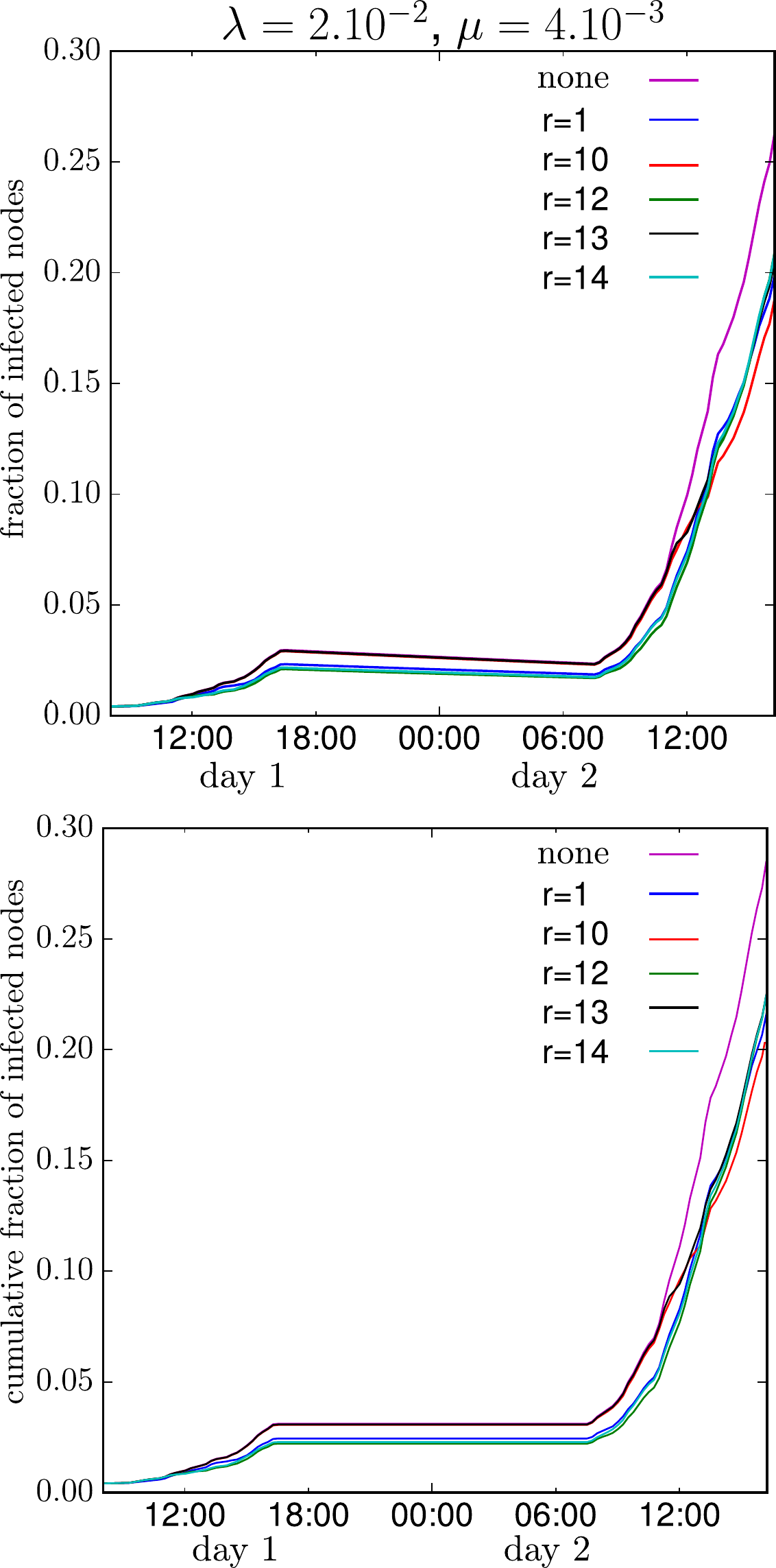}}
    \subfigure[]{\includegraphics[width=0.48\textwidth]{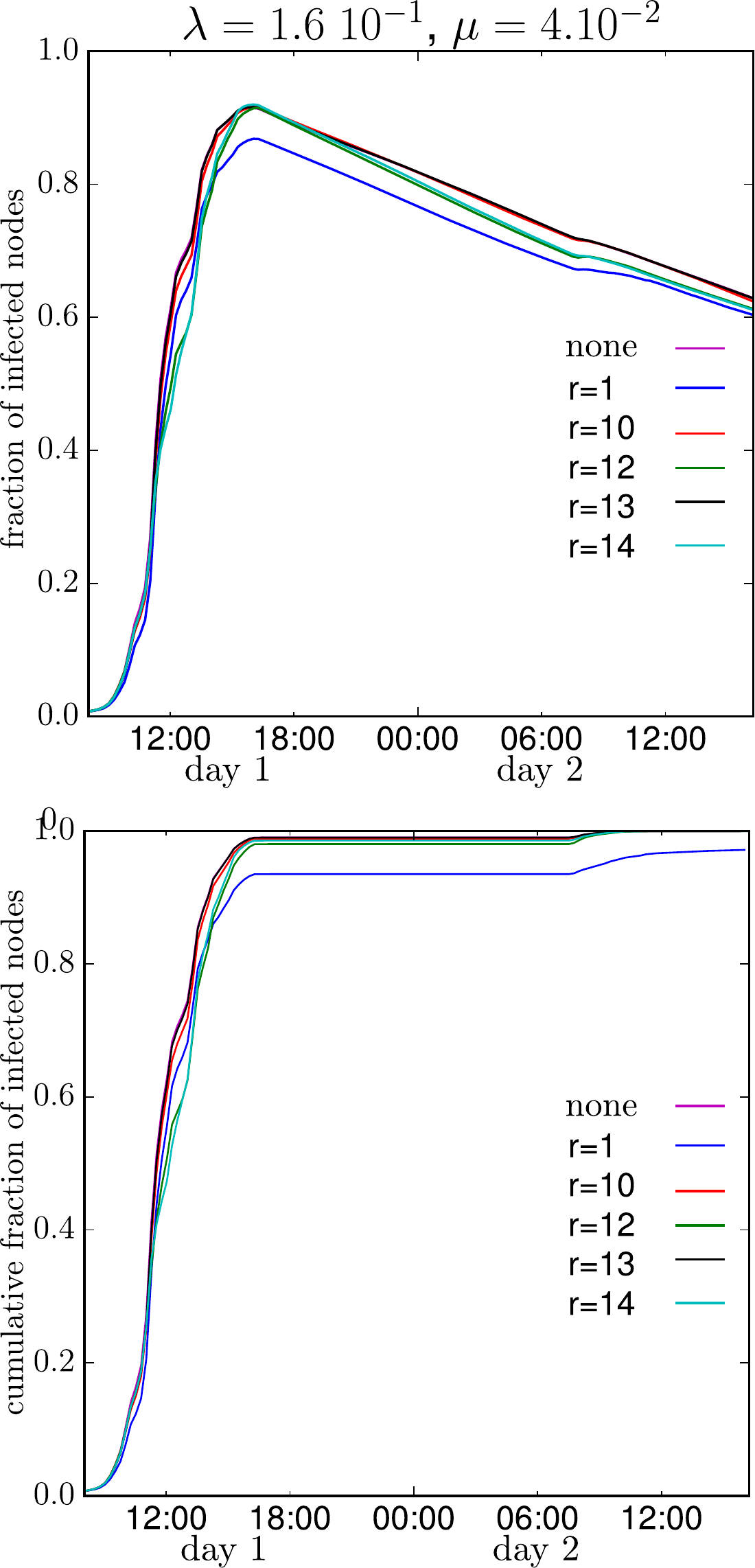}}
    \caption{
   \textbf{ Temporal evolution of the average fraction of nodes in the infected states for SIR processes 
simulated on top of altered networks with a mesostructure $r$ removed
and on top of the whole decomposition $\tilde{\mathcal{T}}$}. The values of $r$ selected here are the same as
in Fig. 4 of the main text,
for  (a) $\lambda=2.10^{-2}$ and $\mu=4.10^{-3}$ and (b)  $\lambda=1.6~10^{-1}$ and $\mu=4. 10^{-2}$. 
The curve marked as \textit{none} corresponds to the spread on the unmodified temporal network $\tilde{\mathcal{T}}$.
}
    \label{fig:evolution}
\end{figure*}

The temporal evolutions of the fraction of infected nodes 
 are displayed in Fig. \ref{fig:evolution} for two sets of $(\lambda,\mu)$ chosen to be in 
two different regimes. In Fig. \ref{fig:evolution}(a), spread and recovery are slow, and
the spread is not over at the end of the dataset.  The removal of  any of the mixing patterns 
thus has  a similar impact at the end of the two days, independently of the fact that their activity occurs on the first or on the second day.
For the parameter values used in Fig. \ref{fig:evolution}(b) on the other hand, the spread is too fast to be mitigated.

\end{document}